# A Molecular Theory of the Nematic-Nematic Phase Transitions in Mesogenic Dimers.

Alexandros G. Vanakaras and Demetri J. Photinos

*Department of Material Science, University of Patras, Patras 26504, Greece.*

**Abstract**

We present a simplified molecular model of mesogenic dimers consisting of two identical uniaxial mesogenic cores separated by a fixed-length spacer and allowed to assume only two, statistically equivalent, conformations which are non-planar and of opposite handedness. In the mean-field approximation, with additive interactions among the mesogenic cores, the model yields up to three positionally disordered phases, one isotropic and two nematic. The low temperature nematic phase ($N_X$) has a local two-fold symmetry axis which is also a direction of molecular polar ordering and is tightly twisted about a macroscopic phase axis. The onset of polar ordering generates spontaneous chiral symmetry breaking, manifested primarily by the twisting of the polar director and the formation of chiral domains of opposite handedness. Within these domains the statistical balance between the two enantiomer conformations is shifted and the principal axes of the ordering tensors of the molecular segments twist at constant tilt angles with the helix axis. Key experimental results on the $N_X$ phase of liquid crystalline dimers are discussed in the light of the theoretical predictions of the model, which are also contrasted with the predictions of the twist-bend nematic model.

*1. Introduction.*

Neat LC compounds which form more than one nematic phase[1–5] are exceptional and bring new insights and challenges to our understanding of the "simplest of all LC phases". A fascinating example is provided by certain types of symmetric liquid crystalline dimers (bimesogens, for short) with odd carbon-number alkyl spacers[6] exhibiting, in addition to the conventional uniaxial nematic phase, a low temperature nematic, termed as Nx[2,7–13]. The latter shows characteristic periodic stripe patterns and rope textures in thin films[2] and an electro-optical response typically found in chiral systems, though the molecules are nonchiral[14–17]. The chiral nature of this phase has been confirmed NMR studies[13,18,19].

Recently, the structure of the Nx phase has received much attention[16,20–24] and, despite the intense investigation, it is still debated. It has often been argued that the Nx should be identified with the theoretically predicted[25] twist-bend nematic phase, $N_{TB}$, which emerges from the instability of the uniaxial nematic phase to spontaneous bend deformations.[9] To date, the analysis and attempted interpretation of a considerable volume of experimental observations has been mostly based on the identification of the bimesogen Nx phase as $N_{TB}$.[7–10,24] However, none of these observations can be considered as direct proof nor does it exclude other possible phase structures. Moreover, the $N_{TB}$ interpretation appears to be in conflict with more recent experimental results.[8,11–13] In this work we present a concrete alternative structural model for the Nx phase, starting from the molecular symmetries and interactions. The molecules are modelled as pairs of uniaxial rods which are rigidly connected in a way that permits only intramolecular rotations which generate distinct conformations, thus capturing minimally the key features of the prototypical symmetric CB-Cn-CB dimer molecules. We demonstrate that, within a certain range of the molecular parameters of the model, a phase transition from a uniaxial achiral nematic phase to a lower temperature nematic phase of locally polar and chiral order is obtained. Based on these results we propose a detailed picture for the molecular organization of the $N_X$ phase which accounts for the occurrence of this phase in some types of dimers and not in others and provides a consistent interpretation for the known experimental results on the structure and the macroscopic properties of the $N_X$ phase.

*2. A toy-model of nematic bimesogens.*

For the symmetric dimer molecule with identical rod-like mesogenic cores we adopt the simplified geometry of figure 1. The directions of the mesogenic rods are denoted by the unit vectors $\mathbf{L}$, $\mathbf{L}'$. The "spacer" vector, connecting the midpoints of the mesogenic rods, is denoted by $\mathbf{d} = d\,\mathbf{z}$ and defines the direction of the molecular $z$-axis. The mesogenic cores form a fixed angle $\beta$ (the molecular "bend" angle) with the spacer vector and the molecule has a twofold symmetry axis perpendicular to $\mathbf{z}$, lets identify it with the molecular $y$ axis. The $z$ axis is also an axis of reflection-rotation by an angle $\alpha$ (the molecular "torsion" angle) which has fixed magnitude but can assume opposite signs $\alpha = \pm|\alpha|$, thus restricting the possible conformations of the dimer molecule to just two. These are mutual enantiomers and will be denoted by a conformation index $s = \pm 1$. The conformations are taken to have intrinsically equal weights, thus rendering the molecule statistically achiral.

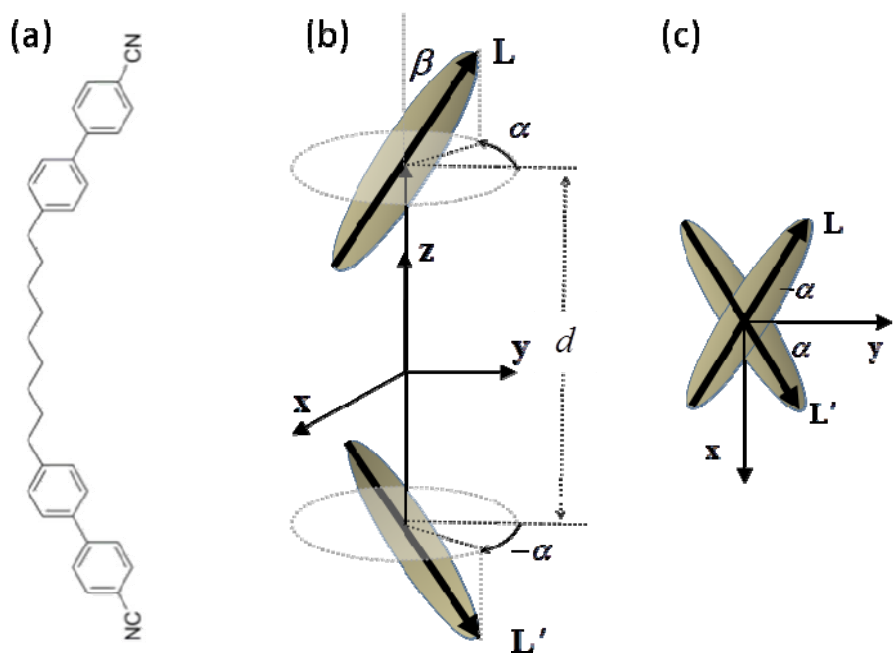

**Figure 1.** Molecular structure of BC-C9-CB (a) and molecular geometry of idealized dimer, side(b) and top (c) view, with the labelling of the molecular axes and mesogenic units..

Despite the extreme simplification of the molecular structure, this minimal representation maintains the crucial features of the actual dimer molecules, namely that (i) the distance between the two mesogenic units cannot exceed an upper bound (the extended spacer length) (ii) the orientations of the mesogenic cores, relative to each other and to the spacer end-to-end direction, are not evenly distributed but show clear statistical biasing about well-defined values. This is illustrated in figure 2 where the statistical distribution[26] of the molecular structure invariant $\mathbf{L}\cdot\mathbf{L}'$ for the mesogenic dimer CB-C9-CB is shown and compared with the respective sharply defined values of the toy-molecule $\mathbf{L}\cdot\mathbf{L}' = -1 + \sin^2\beta(\cos 2\alpha + 1)$.

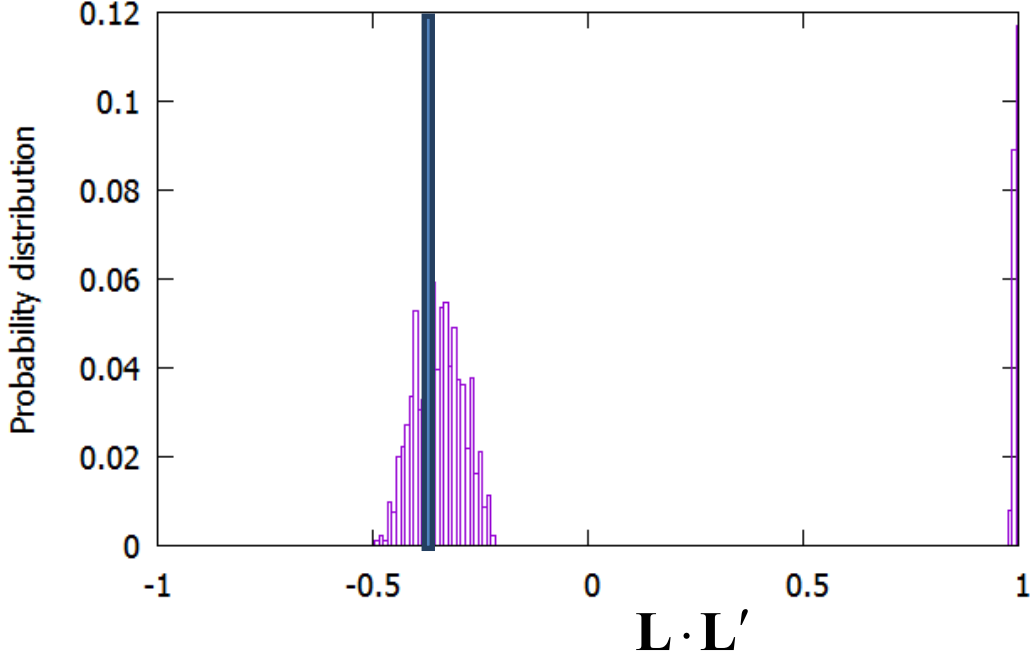

**Figure 2:** Calculated distribution[27] of the relative orientations $\mathbf{L}\cdot\mathbf{L}'$ of the mesogenic cores of the CB-C9-CB dimer molecule in figure 1(a). Shown on the same diagram is the single value (thick line) of $\mathbf{L}\cdot\mathbf{L}'$ obtained for the toy-molecule shown in figures 1(b) and (c) with the angles $\alpha = 30°$ and $\beta = 30°$.

Keeping the formulation of the intermolecular interactions at an analogous level of simplicity, a pair of molecules, 1 and 2, is assumed to interact via pair a potential $V_{1,2}$ that is made up additively of terms involving only the four intermolecular pairs of mesogenic core segments in the two dimers: $V_{1,2}(\mathbf{R}_{1,2},\omega_1,\omega_2,s_1,s_2) = v_{L_1,L_2} + v_{L_1,L_2'} + v_{L_1',L_2} + v_{L_1',L_2'}$, where $\mathbf{R}_{1,2}$ is the intermolecular vector, connecting the origins of the molecular frames of dimers 1 and 2, in conformations $s_1, s_2$, and orientations $\omega_1, \omega_2$ of their respective molecular frames in the macroscopic frame of axes $X, Y, Z$. The core-core potentials $v_{L_1,L_2} = v_{L_1,L_2}(\mathbf{R}_{L_1,L_2}, \mathbf{L}_1, \mathbf{L}_2)$, depend on the intermolecular vector $\mathbf{R}_{L_1,L_2}$ connecting the mid-points of the core segments and on the unit vectors $\mathbf{L}_1, \mathbf{L}_2$ specifying the directions of these segments in the macroscopic frame.

The assumed molecular symmetry can lead to three types of positionally disordered fluid phases, to which we hereafter restrict our attention. These are the isotropic fluid phase (I), the uniaxial nematic phase (N) and a nematic phase with possible transverse polar ordering, and therefore locally biaxial arrangement of the molecules. The latter phase, which will be tentatively termed as $N_X$, could in principle present chiral ordering as well, due to the existence of chiral molecular conformations, and to the overall bent molecular shape . In any case, this phase should present at least a local two-fold symmetry axis associated with the two-fold symmetry of both conformations accessible to the constituent molecules. Accordingly, there is one phase director, denoted by the unit vector $\mathbf{m}$, which is a local twofold symmetry axis of the $N_X$ phase, thus defining the only possible direction of polar molecular ordering in the phase. Of course, being a director of the phase, $\mathbf{m}$ is a principal axis of any tensor quantity in that phase. As the free energy calculations show below, the thermodynamically most stable spatial configuration of the $\mathbf{m}$ director is a helical twisting of constant pitch $p_h$ about a fixed axis perpendicular to $\mathbf{m}$. This helix axis, which is an effective full rotational symmetry axis for any physical property of the phase that is averaged over regions of macroscopic dimensions much larger than the helical pitch, will be denoted by the unit vector $\mathbf{n}_h$. In the conventional uniaxial nematc phase, N, $\mathbf{n}_h$ reduces to the usual director $\mathbf{n}$, the unique, inversion invariant ($\mathbf{n} \Leftrightarrow -\mathbf{n}$), full rotational symmetry axis of the phase. To complete the set of local phase axes for the description the $N_X$ phase, we define a third unit vector $\mathbf{l}_h$, perpendicular to $\mathbf{m}$ and $\mathbf{n}_h$. With this identification of local phase axes, the simultaneous sign reversal $\{\mathbf{n}_h, \mathbf{l}_h\} \Rightarrow \{-\mathbf{n}_h, -\mathbf{l}_h\}$ is a symmetry operation of the phase, while the sign reversal $\mathbf{m} \Rightarrow -\mathbf{m}$ is not necessarily a symmetry operation, thus allowing for the description of transversely polar phases. Obviously, these local symmetries are compatible with a twist of the two-fold (and possibly polar) axis $\mathbf{m}$ about the helix axis $\mathbf{n}_h$. Furthermore, the spontaneous breaking of orientational symmetry which distinguishes the $N_X$ phase from N, dictates that the broken symmetries of the N phase will be preserved globally in the less symmetric $N_X$ phase. Accordingly, full rotational symmetry about an axis and inversion of that axis should be present as global symmetries in the $N_X$ phase. These requirements have direct implications on the possible spatial configurations of the local axes $\mathbf{n}_h, \mathbf{l}_h, \mathbf{m}$: Choosing the frame of macroscopic, phase fixed, axes $X,Y,Z$ such that the $Z$ axis is along the helical axis $\mathbf{n}_h$, and taking into account the macroscopic equivalence of all the positions and directions in the $XY$ plane, we have for the spatial dependence the two transverse phase axes the following possibility $\mathbf{l}_h(Z) = \mathbf{X}\cos\varphi(Z) + \mathbf{Y}\sin\varphi(Z)$ and $\mathbf{m}(Z) = -\mathbf{X}\sin\varphi(Z) + \mathbf{Y}\cos\varphi(Z)$. The equivalence of all the positions along the $Z$ axis implies a constant derivative $d\varphi(Z)/dZ = k$, from which it follows that $\varphi(Z) = kZ + \varphi_0$, i.e. twisting at constant pitch $p_h = 2\pi/k$. The macroscopic equivalence of the $Z$ and $-Z$ directions implies the equivalence of both signs for the twist wave number, $\pm k$. The phase angle $\varphi_0$ is arbitrary and accounts for the arbitrariness in the choice of the $X$ and $Y$ directions and of the origin of the macroscopic frame $X,Y,Z$. Obviously, the uniformly aligned (untwisted) biaxial, and possibly polar, nematic phase is included as in this description and corresponds to a vanishing value $k$.

In the $N_X$ phase, the entire molecule has an orientation-conformation probability $f(s,\omega,Z)$ that refers to the local axes $\mathbf{n}_h, \mathbf{l}_h(Z), \mathbf{m}(Z)$ at the $Z$ coordinate of the origin of the molecular frame of axes. The orientation of the molecular frame $x,y,z$, relative to the macroscopic $X,Y,Z$ is denoted by $\omega$ and the positional distribution of the molecular frame origins is taken to be uniform (i.e. nematic).

To describe phase stability and transitions among the possible positionally uniform phases of the system, we formulate the free energy starting from the average interaction (the "potential of mean torque") experienced by a dimer molecule from all the other molecules in a sample of $N$ molecules of uniform molecular density $\rho(=N/\text{sample volume})$ at temperature $T$. This is obtained by averaging $V_{1,2}$ over all the positions ($\vec{R}_2$), orientations ($\omega_2$) and conformations ($s_2$) of molecule 2.

$$\bar{V}(s_1,\omega_1,Z_1) = \left(\frac{\rho}{2}\right)\sum_{s_2}\int d\vec{R}_2 \int d\omega_2\, g_{1,2} V_{1,2} f(s_2,\omega_2,Z_2) \tag{1}$$

Here $g_{1,2} = g_{1,2}(\mathbf{R}_{1,2},\omega_1,\omega_2,s_1,s_2)$ stands for the pair correlation function of dimers 1 and 2.

To proceed with the determination of the potential of mean torque, we follow closely the Maier-Saupe (M-S) approach,[28,29] with the crucial difference that here the positional and orientational averaging is not done independently among the mesogenic units, as in the M-S model of nematics, but among permanently jointed pairs of such units, forming the individual dimer molecules. Accordingly, we make the drastically simplifying approximation of using for each of the four segment-segment terms in $V_{1,2}$ a factorized form in which the dependence on the relative orientations is separated from the dependence on the relative distance of the segments[28,29], that is

$$g^{(2)}(\mathbf{R}_{1,2},\omega_1,\omega_2,s_1,s_2)v_{L_1,L_2}(\mathbf{R}_{L_1,L_2},\mathbf{L}_1,\mathbf{L}_2) \approx u(R_{L_1,L_2})P_2(\mathbf{L}_1 \cdot \mathbf{L}_2). \quad (2)$$

Here $P_2(\mathbf{L}_1 \cdot \mathbf{L}_2)$ denotes the second Legendre polynomial of the angle formed by the intermolecular pair of mesogenic rods $\mathbf{L}_1, \mathbf{L}_2$. The radial function $u(R_{L_1,L_2})$ is understood to vanish at short segment-segment distances $R_{L_1,L_2}$ (as molecular overlaps are inhibited by the pair correlation) and at large distances (finite range interactions).

Thus, from each of the additive pair terms, say the $v_{L_1,L_2}$ term, we have the contribution:

$$\overline{v}_{L_1}(s_1,\omega_1,Z_1) = \left(\frac{\rho}{2}\right)\sum_{s_2}\int d\vec{R}_2 \int d\omega_2 \, u(R_{L_1,L_2}) f_{s_2}(s_2,\omega_2,Z_2) P_2(\mathbf{L}_1 \cdot \mathbf{L}_2). \quad (3)$$

To carry out the integrations in the above expression we introduce the Fourier representation

$$u(R) = \int d\vec{q} \, u'(q) e^{i\vec{q}\cdot\vec{R}} \quad \text{with} \quad u'(q) = \frac{1}{2\pi^2 q}\int_0^\infty u(R)(\sin qR) R \, dR, \quad (4)$$

and we use the relations

$$\begin{aligned}\mathbf{m}(Z_2) &= -\mathbf{l}_h(Z_1)\sin k(Z_2 - Z_1) + \mathbf{m}(Z_1)\cos k(Z_2 - Z_1) \\ \mathbf{l}_h(Z_2) &= \mathbf{l}_h(Z_1)\cos k(Z_2 - Z_1) + \mathbf{m}(Z_1)\sin k(Z_2 - Z_1)\end{aligned}, \quad (5)$$

and

$$\vec{R}_{L_1,L_2} = \vec{R}_2 - \vec{R}_1 + \vec{r}_{L_2} - \vec{r}_{L_1}, \quad (6)$$

where the molecular vector $\vec{r}_{L_1}$ connects the molecular centre of molecule 1 with the centre of the segment $L_1$ and is therefore equal to $\vec{r}_{L_1} = \mathbf{z}_1(d/2)$ (respectively $\vec{r}'_{L_1} = -\mathbf{z}_1(d/2)$).

We then obtain from eq(3), on adding the contributions from the four possible intermolecular mesogenic pairs, the following expression for the potential of mean torque of the dimer molecule in the $N_X$ phase:

$$-\overline{V}(s,\omega)/k_B T = \varepsilon \left( W_0(s,\omega)\langle W_0 \rangle + u_1 W_1(s,\omega)\langle W_1 \rangle + u_2 W_2(s,\omega)\langle W_2 \rangle \right), \tag{7}$$

with the angular brackets denoting ensemble average of any conformation /orientation /position-dependent molecular quantity $Q(s,\omega,Z)$ according to $\langle Q \rangle \equiv (1/2)\sum_s \int d\omega f(s,\omega,Z) \times Q(s,\omega,Z)$ and with the following analytic expressions for the orientation-conformation dependent molecular quantities $W_i$, $i = 0,1,2$:

$$W_0(s,\omega) = \left(\frac{3}{2}\cos^2\beta - \frac{1}{2}\right)\left(\frac{3}{2}(\mathbf{z}\cdot\mathbf{n}_h)^2 - \frac{1}{2}\right) - \frac{3}{4}\sin^2\beta\cos 2\alpha\left((\mathbf{x}\cdot\mathbf{n}_h)^2 - (\mathbf{y}\cdot\mathbf{n}_h)^2\right)$$
$$-\frac{3}{2}\sin 2\beta \sin\alpha (\mathbf{z}\cdot\mathbf{n}_h)(\mathbf{x}\cdot\mathbf{n}_h) \tag{8a}$$

$$W_1(s,\omega) = \begin{pmatrix} \left(\frac{3}{2}\cos^2\beta - \frac{1}{2}\right)(\mathbf{z}\cdot\mathbf{n}_h)(\mathbf{z}\cdot\mathbf{n}_h) \\ +\frac{1}{2}\sin^2\beta\cos 2\alpha\left((\mathbf{y}\cdot\mathbf{n}_h)(\mathbf{y}\cdot\mathbf{l}) - (\mathbf{x}\cdot\mathbf{n}_h)(\mathbf{x}\cdot\mathbf{l})\right) \\ -\frac{1}{2}\sin 2\beta \sin\alpha \left((\mathbf{z}\cdot\mathbf{l})(\mathbf{x}\cdot\mathbf{n}_h) + (\mathbf{z}\cdot\mathbf{n}_h)(\mathbf{x}\cdot\mathbf{l})\right) \end{pmatrix} \cos k^*(\mathbf{z}\cdot\mathbf{n}_h)$$
$$+\frac{1}{2}\begin{pmatrix} \sin 2\beta \cos\alpha \begin{pmatrix} (\mathbf{z}\cdot\mathbf{m})(\mathbf{y}\cdot\mathbf{n}_h) \\ +(\mathbf{y}\cdot\mathbf{m})(\mathbf{z}\cdot\mathbf{n}_h) \end{pmatrix} \\ -\sin^2\beta\sin 2\alpha \begin{pmatrix} (\mathbf{y}\cdot\mathbf{n}_h)(\mathbf{x}\cdot\mathbf{m}) \\ +(\mathbf{x}\cdot\mathbf{n}_h)(\mathbf{y}\cdot\mathbf{m}) \end{pmatrix} \end{pmatrix} \sin k^*(\mathbf{z}\cdot\mathbf{n}_h) \tag{8b}$$

and

$$W_2(s,\omega) = \begin{pmatrix} \left(\dfrac{3}{2}\cos^2\beta - \dfrac{1}{2}\right)\left((\mathbf{z}\cdot\mathbf{m})^2 - (\mathbf{z}\cdot\mathbf{l}_h)^2\right) \\ +\dfrac{1}{2}\sin^2\beta\cos 2\alpha\begin{pmatrix}(\mathbf{y}\cdot\mathbf{m})^2 - (\mathbf{y}\cdot\mathbf{l}_h)^2 \\ -(\mathbf{x}\cdot\mathbf{m})^2 + (\mathbf{x}\cdot\mathbf{l}_h)^2\end{pmatrix} \\ +\sin 2\beta\sin\alpha\left((\mathbf{z}\cdot\mathbf{l}_h)(\mathbf{x}\cdot\mathbf{l}_h) - (\mathbf{z}\cdot\mathbf{m})(\mathbf{x}\cdot\mathbf{m})\right) \end{pmatrix} \times \cos 2k^*(\mathbf{z}\cdot\mathbf{n}_h)$$

$$+ \begin{pmatrix} \sin 2\beta\cos\alpha\begin{pmatrix}(\mathbf{z}\cdot\mathbf{m})(\mathbf{y}\cdot\mathbf{l}_h) \\ +(\mathbf{y}\cdot\mathbf{m})(\mathbf{z}\cdot\mathbf{l}_h)\end{pmatrix} \\ -\sin^2\beta\sin 2\alpha\begin{pmatrix}(\mathbf{y}\cdot\mathbf{l}_h)(\mathbf{x}\cdot\mathbf{m}) \\ +(\mathbf{x}\cdot\mathbf{l}_h)(\mathbf{y}\cdot\mathbf{m})\end{pmatrix} \end{pmatrix} \times \sin 2k^*(\mathbf{z}\cdot\mathbf{n}_h) \qquad (8c)$$

In these expressions, the dimensionless inverse pitch is defined as $k^* \equiv kd/2 = \pi d/p_h$. The overall strength parameter appearing in the potential of mean torque is given by $\varepsilon \equiv -4(2\pi)^3 \rho u'(0)/k_B T$ and the relative strength parameters of the first and second "twist harmonic" contributions in eq (7) are $u_1 \equiv 3u'(k)/u'(0)$; $u_2 \equiv 3u'(2k)/4u'(0)$. The normalized orientation-conformation distribution function is expressed in terms of the potential of mean torque as

$$f(s,\omega) = \frac{\exp\left(-\bar{V}(s,\omega)/k_B T\right)}{\zeta\left(\langle W_0\rangle, \langle W_1\rangle, \langle W_2\rangle\right)} \quad , \qquad (9)$$

with $\zeta \equiv 2/\sum_s \int d\omega \exp\left(-\bar{V}(s,\omega)/k_B T\right)$, and the respective free energy of the ensemble[28] is

$$F/k_B T = \frac{1}{2}\varepsilon\left(\langle W_0\rangle^2 + u_1\langle W_1\rangle^2 + u_2\langle W_2\rangle^2\right) - \ln\zeta\left(\langle W_0\rangle, \langle W_1\rangle, \langle W_2\rangle\right) \quad . \qquad (10)$$

The three composite order parameters $\langle W_0\rangle, \langle W_1\rangle, \langle W_2\rangle$ together with the inverse pitch parameter $k^*$ are determined self-consistently by minimizing the free energy functional of eq (10). The following types of solutions are considered, corresponding to the following fluid phases: (i) Isotropic, I, with $\langle W_0\rangle = \langle W_1\rangle = \langle W_2\rangle = 0$; (ii) uniaxial nematic, N, with $\langle W_0\rangle \neq 0, \langle W_1\rangle = \langle W_2\rangle = 0$; (iii) biaxial apolar nematic, with $\langle W_0\rangle \neq 0, \langle W_1\rangle \neq 0, \langle W_2\rangle = 0$, with uniformly aligned ($k^* = 0$) or twisted ($k^* \neq 0$) transverse director **m**; (iv) transversely polar (and therefore biaxial) nematic, with $\langle W_0 \rangle \neq 0, \langle W_1 \rangle \neq 0, \langle W_2 \rangle \neq 0$, with uniformly aligned ($k^* = 0$) or twisted ($k^* = 0$) **m**. The N$_X$ phase is identified with the latter case, i.e. as a transversely polar (and therefore biaxial) nematic phase with twisted transverse director about a uniformly oriented helical axis $\mathbf{n}_h$. Our calculations yield only the I, N and N$_X$ solutions, with the latter, when obtained, being always more stable thermodynamically than the N solution.

As the free energy is invariant with respect to the simultaneous reversal of the molecular torsion angle *α* (i.e. replacing each conformation by its enantiomer) and the reversal of the secondary director twisting sense, i.e. the reversal of the macroscopic chirality, domains of opposite chirality are equally likely to be present in an unbiased macroscopic sample of the N$_X$ phase. Also, as expected, the free energy shows no dependence on the phase angle *φ$_0$* associated with the arbitrariness in the choice of the origin and the transverse directions of the macroscopic frame $X, Y, Z$. Accordingly, the variation of this phase angle from one domain to the other is unrestricted, aside from surface terms at the interface with neighboring domains.

Results on the relative thermodynamic stability of these phases and on the respective phase transitions are presented in the next section. It should be kept in mind that these results refer to unperturbed macroscopic monodomain ensembles at fixed density and temperature.

### *3. Phase transition and order parameter calculations.*
Here we restrict our consideration to a range of the molecular angles *α* (the "torsion" angle) and $\beta$ (the "bend" angle) that are relevant to the molecular architecture of the odd-spacer mesogenic dimers. Obviously, for very small bend angles the behaviour of the system tends to that of typical uniaxial rod nematics while for bend angles near *π/2* and torsions either close to 0 or to $\pi/2$, the behavior tends to that of plate-like nematics. Accordingly we will consider, for both angles, values not very far from $\pi/4$. Furthermore, to have a concrete analysis of the phase transition trends and of the order parameter profiles, we have employed, in line with the simplicity of the molecular model, a very simple form for the radial part of the effective mesogenic core interaction in eq (4), namely $u(R) = -u_{atr}$ in the range $R_0 < R < R_a$ and $u(R) = 0$ otherwise. This leads to the following functional dependence of $u'(q)$

$$u'(q) = \frac{1}{2\pi^2} \times \frac{u_{atr}}{q^3}\left[\left(\sin qR_0 - qR_0 \cos qR_0\right) - \left(\sin qR_a - qR_a \cos qR_a\right)\right] , \quad (11)$$

which conveys in a very simple form the main physical features of the interactions. Calculations with different forms indicate that the qualitative inferences of the model do not depend critically on the details of the functional form used for $u'(q)$.

We now turn to the dependence of the possible phase sequences and respective transition temperatures on the molecular bend and twist angles $\beta$ and $\alpha$. The temperature dependence of the potential of mean torque in eq(7) is controlled by the strength parameter $\varepsilon$, which also controls the temperature dependence of the composite order parameters $\langle W_0 \rangle, \langle W_1 \rangle, \langle W_2 \rangle$ and of the helical pitch, through the free energy minimisation conditions. Accordingly, the parameter $\varepsilon$ can be regarded as an external parameter by which the effective inverse temperature of the system is specified. For convenience in the presentation of the phase diagrams we use a reference value of $\varepsilon$, specifically the value $\varepsilon_{N-I}(\beta = 0)$ at the nematic to isotropic transition for the limiting case of $\beta = 0$, corresponding to a two-rod linear dimer. We then express the effective inverse temperature of the system for any combination of the $\beta$, $\alpha$ angles by means of the dimensionless parameter $\varepsilon^* = \varepsilon(\beta,\alpha) / \varepsilon_{N-I}(\beta = 0)$.

The solution of the self-consistency equations for the composite order parameters $\langle W_0 \rangle, \langle W_1 \rangle, \langle W_2 \rangle$ and the pitch $k^*$, together with the evaluation of the free energy for each of the possible solutions show that (i) for relatively small bend angle $\beta$ and large torsion angle $\alpha$, a first-order transition from the Isotropic (I) to the uniaxial nematic (N) phase is obtained at a temperature $T_{N-I}$ and on further lowering the temperature, a second-order phase transition is found from the nematic N to a transversely polar and twisted nematic phase ($N_X$) and (ii) for larger bend angles $\beta$ and relatively small torsion angles $\alpha$, a direct first order I- $N_X$ phase transition is obtained at low temperatures. Representative results of the calculations are shown in the phase diagrams of figure (3) for variable molecular bend angle at fixed torsion angle and *vice versa*. In both cases, a critical value of the variable angle is obtained, at which the three phases coexist and beyond which the uniaxial nematic phase, N, is eliminated from the phase sequence. The topology of these diagrams remains essentially unchanged for different values of the respective fixed angle ($\beta$ or $\alpha$) within a range of $\pm 15°$ about the chosen value of $30°$ in the two diagrams of figure 3. A generic feature of the $\beta$-dependence of the phase diagrams is the rapid lowering of the N-$N_X$ transition temperature with decreasing bend-angle, suggesting that the latter is the primary molecular feature controlling the stabilization of the $N_X$ relative to other competing phases (smectic, etc.) which are accessible to the bimesogens at low temperatures.

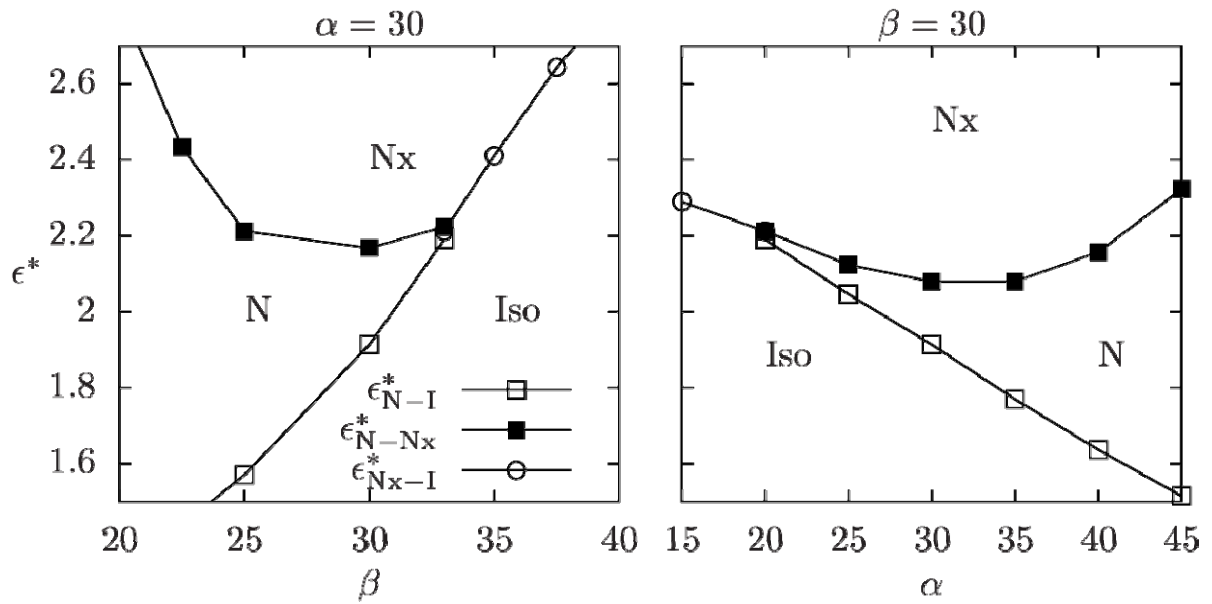

**Figure 3.** Calculated phase diagrams for fixed molecular torsion angle $\alpha = 30°$ and variable molecular bend angle $\beta$ (left) and for fixed $\beta = 30°$ and variable $\alpha$ (right). The dimensionless parameter in the vertical axis is the effective inverse temperature $\varepsilon^*$. The lines dividing the different phase regions correspond to the respective phase transition temperatures. The results are obtained for the molecular interaction range parameter values $R_0 / d = 1$ and $R_a / R_0 = 1.5$, for which, $\varepsilon_{N-I}(\beta = 0) = 4.52$.

Results on the temperature dependence of the order parameters are shown in figure 4, with the molecular bend and torsion angles fixed at $\alpha = \beta = 30°$. Of the many possible order parameter combinations describing the $N_X$ phase of the system, we have chosen the following, mainly due to their more or less direct accessibility to experimental determination: (i) Orientational order parameters in the frame of the axes $\mathbf{n}_h$, $\mathbf{l}_h$, $\mathbf{m}$. These include the second rank order parameters of the mesogenic cores $S_h^{(L)} = \left\langle \frac{3}{2}(\mathbf{L} \cdot \mathbf{n}_h)^2 - \frac{1}{2} \right\rangle$, measuring the extent of alignment of the mesogenic units along the helix axis $\mathbf{n}_h$, the order parameter $\Delta_h^{(L)} = \left\langle (\mathbf{L} \cdot \mathbf{m})^2 - (\mathbf{L} \cdot \mathbf{l}_h)^2 \right\rangle$, measuring the local biaxiality of the mesogenic core orientational ordering, and the respective local eccentricity parameter $\delta_h^{(L)} = \left\langle (\mathbf{L} \cdot \mathbf{n}_h)(\mathbf{L} \cdot \mathbf{l}_h) \right\rangle$. They also include the first rank order parameter $P = \langle \mathbf{y} \cdot \mathbf{m} \rangle$, measuring the polar ordering of the molecular $C_2$ axis along the director $\mathbf{m}$. The polar ordering of the mesogenic units, which would produce the spontaneous electric polarisation $\mathbf{P}^{(L)}$, along $\mathbf{m}$, if these units were endowed with longitudinal dipole moments, is given by the order parameter $P^{(L)} = \langle \mathbf{L} \cdot \mathbf{m} \rangle = P \cos\alpha \sin\beta$. (ii) The conformational order parameter $c = \langle s \rangle = \langle \sin\alpha \rangle / |\sin\alpha|$, which provides a measure of the induced imbalance between the two enantiomeric conformations, in other words, the molecular chirality symmetry breaking and (iii) the dimensionless inverse pitch $k^*$ of the polar director twisting. Clearly, only the $S_h$ order parameter survives in the uniaxial nematic phase, where it reduces to the nematic order parameter $S$. In the $N_X$ phase, the other parameters, $\Delta_h^{(L)}, \delta_h^{(L)}, P, c$ and $k^*$, acquire non vanishing values, with both signs of the polarity $P$, chirality $c$ and twisting sense $k^*$ being thermodynamically equivalent. Therefore domains of opposite polarity and handedness can coexist in different regions of an unbiased macroscopic sample.

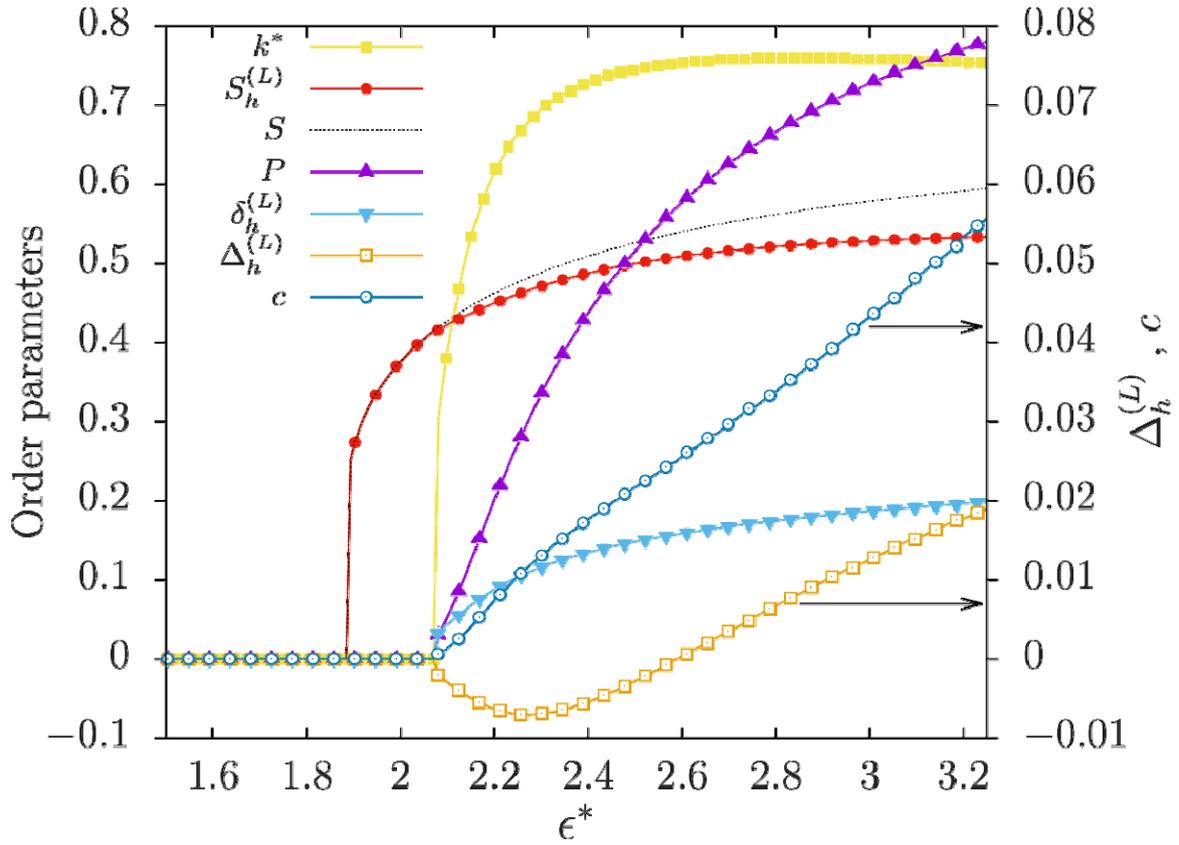

**Figure 4.** Calculated temperature dependence of the second rank orientational order parameters of the mesogenic core segments, $S_h^{(L)}$, $\Delta_h^{(L)}$ and $\delta_h^{(L)}$, defined in the main text, the polarity order parameter $P$, the induced molecular chiral asymmetry parameter $c$ and the dimensionless inverse pitch $k^*$ of the twisting of the transverse polar director $\mathbf{m}$. The dotted line represents the calculated extrapolation of the nematic order parameter $S$ below the transition temperature, where the uniaxial nematic phase becomes thermodynamically unstable. Note the different scale for $\Delta_h^{(L)}$, $c$. All the results are obtained for fixed values of the molecular angles $\alpha = \beta = 30°$ and of the interaction range parameters $R_0/d$, $R_a/d$ as in figure 3.

The transition from the isotropic, I, to the uniaxial nematic phase, N, shows the typical features of the usual, weakly first order, N-I transition. Regarding the transition from the uniaxial nematic N to the $N_X$ phase, it is apparent from the temperature profiles in figure 4, that below the transition temperature all the parameters grow continuously starting from their values in the N phase but with a change in the slope. Notably, the order parameter $S_h$ grows slower, with decreasing temperature, than the extrapolated nematic phase order parameter $S$, a trend that has been observed experimentally in several studies[8,13,18,19]. Also, the biaxiality parameter $\Delta_h$ remains very small in absolute value (<0.02) throughout the $N_X$ temperature range, and appears to undergo a change of sign with decreasing temperature. The eccentricity parameter $\delta_h$ acquires substantial magnitude with decreasing temperature in the $N_X$ phase, indicating substantial deviation of the principal axes of the mesogenic unit ordering tensor from the $\mathbf{n}_h, \mathbf{l}_h$ directions. We shall return to this point in section 4.4. Interestingly, the calculated conformational shift, quantified by the induced molecular chirality parameter $c$, remains rather small. In sharp contrast, the polar order parameter $P$ is large and, unlike $S_h$, does not show saturation tendency with decreasing temperature. Finally, the dimensionless inverse pitch $k^*$ shows a continuous but very steep increase just below the transition temperature, rapidly reaching a nearly constant value in the rest of the $N_X$ temperature range. This behavior is observed experimentally by different methods, which also determine the pitch to be rather short, on the order of 10nm. The value at which $k^*$ appears to rapidly level of in figure 4 is between 0.7 and 0.8, leading to the calculated value of the pitch $p_h \approx 4d$, i.e. roughly four molecular lengths, in good agreement with the experimental estimates.

The calculated trends shown in the profiles of Figure 4 remain qualitatively unchanged and with minor quantitative variations on changing the fixed values of the molecular angles $\alpha, \beta$ by ±15°. These trends, as well as the topologies of the phase diagrams in figure 3, do not change substantially for moderate variations of the interaction range parameters $R_0, R_a$ about the indicated values. Specifically, increasing the ratio $R_a / R_0$ tends to destabilize the $N_X$ phase in favour of the N phase. Increasing $R_a / d$, at fixed $R_a / R_0$, leads to increasing values of the pitch (decreasing $k^*$) and destabilization of the $N_X$ phase. On the other hand, for extremely short ranged interactions, $R_a \ll d$, the free energy of the $N_X$ phase may exhibit a second minimum with respect to $k^*$. The solution corresponding to the high $k^*$ minimum is always more stable at low temperatures. Increasing the temperature is found to lead either to the merging of the two minima into a single shallow minimum, producing a smooth crossover from the high to the low $k^*$ phase, or to the eventual stabilization of the large $k^*$ solution, thus producing a first order phase transition from a tightly twisted $N_X$ phase to loosely twisted one on heating. A detailed account of this interesting behavior is given elsewhere.[30]

## *4. Discussion.*

The results presented in the previous section, being derived directly from an explicit molecular model of the characteristic bimesogen structure and showing remarkable robustness with respect to variations of the molecular parameters, offer clear insights into the ordering in the $N_X$ phase and the relation of its distinguishing properties to the key features of the molecular structure. The main implications of the theoretical results and their testing against experiment are discussed below.

### *4.1 The nature of the N-$N_X$ phase transition.*

The transition from the uniaxial nematic to an unperturbed macroscopic monodomain $N_X$ phase is a continuous transition from an apolar to a polar and twisted phase. This orientational order-disorder transition is driven by polar molecular packing in the direction of the molecular $C_2$ (the $y$-molecular axis in figure 1, see also figure 5). The primary orientational order parameter in this transition is the transverse spontaneous polarisation $P$. The onset of transverse polar ordering is accompanied by spontaneous symmetry breaking of chirality, which is manifested by the helical twisting of the spontaneous polarization direction **m** about a unique helix axis $\mathbf{n}_h$. The twisting is dictated by molecular packing frustration, resulting from the bend arrangement of the mesogenic units in the bimesogens and emerges as the thermodynamically most stable mode of optimising the mutual alignment of the mesogenic units while maintaining their transversely polar ordering. The appearance of local biaxiality $\Delta_h$ and some deviation $c$ from the balance between the two enantiochiral molecular conformations are secondary consequences of this mechanism. Also, the appearance of the local eccentricity $\delta_h$, together with the suppressed growth of $S_h$ with decreasing temperature, reflect the symmetry reduction from $D_{\infty v}$ in the N phase to the monoclinic $C_2$ in the $N_X$ phase as a result of the onset of transverse polar ordering.

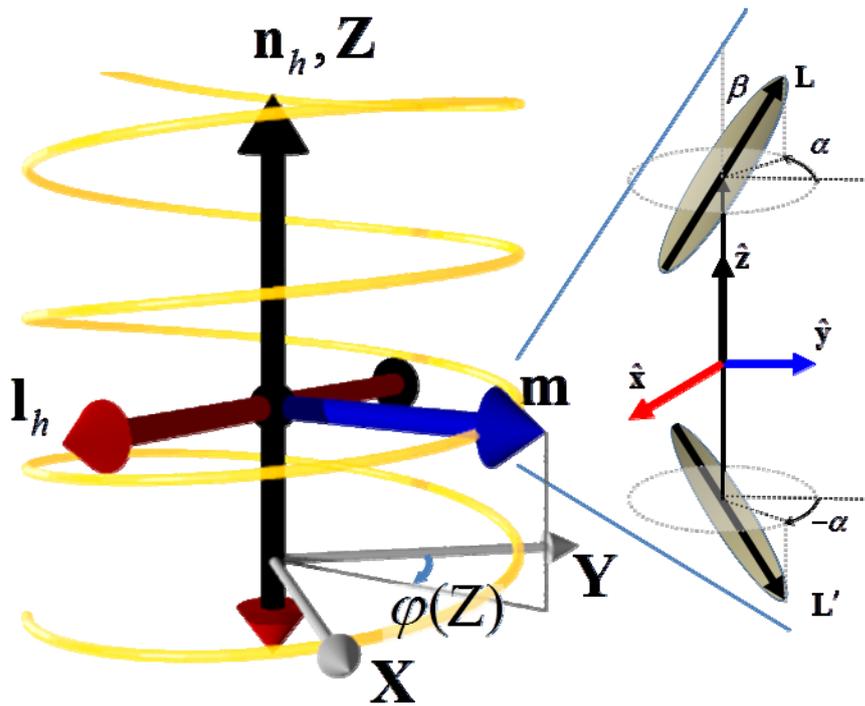

**Figure 5.** Schematic illustration of the structure of the Nx phase, showing the twisting of the polar director **m** and the preferred configuration of the toy-dimer molecules, with the molecular twofold axis **y** tending to order along **m**.

It is important to note that polar ordering is not necessarily generated by electric dipole interactions, as the model molecules are not endowed with permanent dipole moments (despite the fact that the prototype CB-C9-CB dimers in figure 1a have strong terminal dipole moments along the mesogenic unit axis). The polarity in this model originates from the shape of the molecules, which are formed by connecting intrinsically non polar segments into a polar structure. This structure is also statistically achiral, which implies the onset of twisting via a spontaneous symmetry breaking mechanism. The statistical achirality of the individual molecules renders both senses of twisting (corresponding to $\pm k^*$) thermodynamically equivalent. This leads to the formation of enantiochiral domains in a macroscopic unbiased sample and could have significant implications on the nature of the N- $N_X$ transition in such samples. Thus, unless particular restrictions are applied, a monodomain-poly domain transition will be superimposed on the N-$N_X$ monodomain transition and therefore the experimental thermodynamic signature of an unbiased sample will reflect the combined effect of the two transitions. This may alter the strictly continuous character of the N-$N_X$ monodomain transition derived from the model. Furthermore, the thermodynamic equivalence of the two enantiochiral forms of domains implies that a slight chiral bias (chiral dopant, surface anchoring) can have giant chirality enhancement effect on a $N_X$ sample, leading to complete elimination of the domains of the opposite handedness, as observed in recent experiments.[23] Lastly, the local monoclinic symmetry of the $N_X$ phase obtained in these calculations, derives from the $C_2$ symmetry of the molecules. Relaxing the latter symmetry, for example by making the two mesogenic units inequivalent, opens up the possibility of triclinic phase symmetry and deviations from the twisting of the polarisation at right angles to a helical axis. Such possibility will not be further considered here and the discussion will be restricted to the monoclinic local symmetry and twisting of the polar director at right angles to the helix axis.

*4.2 The direct I-$N_X$ transition.*

As illustrated in the phase diagrams of figure 3, the combination of large molecular bend angles $\beta$ with relatively small torsion angles $\alpha$ tend to destabilize the uniaxial nematic phase, thus giving rise to direct transitions from the isotropic to the $N_X$ phase. This is a first order transition, with the order parameters $S_h^{(L)}, \delta_h^{(L)}, k^*$ changing discontinuously across the transition from 0 to nearly their saturation values, the spontaneous polarization $P$ showing a discontinuous jump to large values and further increasing with decreasing temperature and with the conformational shift $c$ and biaxiality $\Delta_h^{(L)}$ showing also discontinuous jumps, albeit of small magnitude. Experimental evidence of this transition has been reported recently.[23] The present model relates the condition of appearance of the direct I-$N_X$ transition to the molecular structural features. Of course, the possibility of this transition being, under the same conditions, pre-empted by a transition to a smectic or crystal phase, cannot be dismissed within the present model, which is restricted to positionally disordered phases only. In any case, the observation of the direct I-$N_X$ transition stresses the nature of $N_X$ as a thermodynamically distinct nematic phase, in contrast to its interpretation as structural deformation of the N phase resulting from an anomaly of its elastic constants. It should also be noted that the spontaneous twisting of the polar director **m** found here is a defining feature of the $N_X$ phase and not merely one of the possible textures that can be exhibited by a biaxial phase.[31]

*4.3 Manifestations of the $N_X$ molecular ordering.*

The model implies strongly polar and chiral molecular ordering in the $N_X$ phase, with tight twisting of the polar direction **m**. Due to this twisting, the polar ordering is averaged out over distances of a few nanometers, and this makes it difficult to directly identify its effects by the conventional electrical/optical methods. On the other hand, the same twisting is associated with the chirality of the molecular order. This has direct manifestations through the periodic stripe patterns[2], the chiral electro-optic response [14–17] and, notably, the NMR spectra of site-labelled bimesogens[13,18,19], because such spectra can provide direct quantitative information on the anisotropically averaged interactions of a molecule with its environment. Specifically, a chiral environment causes the doubling of certain spectral lines which would collapse into a single line in a non-chiral environment.[32] Furthermore, as the frequencies of the spectral lines, whether in a chiral or achiral environment, are related to the orientational order of the respective molecular segments, the spectra allow the quantification of the orientational molecular ordering and of the chiral asymmetry, if present, in such ordering. Due to the extreme simplification of the molecular structure in the present model, it cannot be used directly to extract quantitative inferences from the spectra of actual bimesoges. It can however be used to elucidate certain crucial qualitative features of the NMR spectra in the $N_X$ phase. For example, the model provides for the quadrupolar frequencies $\Delta v_Q$ of a dueteriated site that is rigidly attached to the mesogenic units the following expressions

$$\left[\Delta v_Q\right]_{H\|Z} \sim S_h^{(L)} \quad ; \quad \left[\Delta v_Q\right]_{H\|Y} \sim -\frac{1}{2}S_h^{(L)} + \frac{3}{4}\Delta_h^{(L)}\cos 2\varphi(Z) \, , \qquad (12)$$

for the helical axis $\mathbf{n}_h$ oriented parallel ($H\|Z$) or perpendicular ($H\|Y$) to the spectrometer magnetic field. According to the second expression, $\Delta v_Q$ generally depends on the $Z$ coordinate and therefore the spectrum obtained from the entire sample should be a superposition of spectral lines from different parts of the sample along the helix axis, yielding a broadened spectral line-shape. However, as the value biaxiality order parameter $\Delta_h$ is persistently found to be two orders of magnitude smaller than $S_h$, the model predicts that, to within 1-2%, the frequencies $\left[\Delta v_Q\right]_{H\|Y}$ of the "π/2 flipped" spectra are Z-independent and equal to half the aligned spectrum frequencies $\left[\Delta \nu_Q\right]_{H \| Z}$. This is precisely the experimental result reported for CB-C9-CB.[13]

As the above NMR method cannot provide direct information on the polar ordering of the $N_X$ phase, we briefly discuss electrical and optical methods in the context of polar ordering investigation. For electrically polar mesogenic units the polar ordering would give rise to a strong spontaneous electric polarisation which couples linearly to an applied electric field. However, due again to the tight twisting of the spontaneous polarization vector, rather high field strengths would be required to produce appreciable electro-optic effects. Specifically, with the helical pitch being two orders of magnitude below optical wavelengths, a uniformly aligned $N_X$ sample will appear optically uniaxial, with the optical axis coinciding with the helix axis $\mathbf{n}_h \| Z$. Applying an electric field in the transverse direction, the helical structure is deformed and this shifts the effective optical axis away from the $Z$ axis. In addition to the direct linear coupling with the spontaneous polarization, contributions to the distortion of the helical structure are possible from the usual bilinear dielectric coupling and from the flexoelectric coupling, in analogy with the distortion of the helix in conventional chiral nematics[33]. The distortions are estimated to be rather small and, in general, it is not straightforward to sort out the contributions from the different couplings.[30] Lastly, the birefringence associated with effective optical uniaxiality of the $N_X$ phase is observed experimentally to decrease with decreasing temperature and, for certain dimer compounds, in the low temperature range of the N phase as well.[7] This apparently anomalous temperature dependence of the birefringence seems to follow in some dimer compounds the temperature dependence of the order parameter $S_h^{(L)}$ across the N- $N_X$ transition[7] or to show a steeper decrease in others.[8] The latter case could be due to a change in the values of the effective molecular polarizability components as a result of the substantially different averaging of the intermolecular interactions in the N and $N_X$ phases, thus reflecting, albeit indirectly, the effects of the strongly polar ordering. Thus, altogether, the conventional optical and electro-optical experimental observations do not seem to provide direct information into the polar ordering of the molecules in the $N_X$ phase.

*4.4 Apparent similarities and important differences from the nematic twist-bend model.*

The considerable values of the eccentricity parameter $\delta_h^{(L)}$ indicate that the helix axis $\mathbf{n}_h$ is not the axis of maximal alignment of the mesogenic segments $\mathbf{L}, \mathbf{L}'$. In other words, it is not a principal axis of their ordering tensor. To determine the principal axis frame of this tensor for a given value of the Z coordinate, i.e. on a given plane perpendicular to the helix axis, we note that the director $\mathbf{m}$, being a symmetry axis, is necessarily one of the three principal axes. The other two, denoted by $\tilde{\mathbf{n}}^{(L)}, \tilde{\mathbf{l}}^{(L)}$, are obtained by diagonalizing the ordering matrix through a rotation of the $\mathbf{n}_h, \mathbf{l}_h$ axes by an angle $\tilde{\theta}^{(L)}$, with $\tan 2\tilde{\theta}^{(L)} = 2\delta_h^{(L)} / (S_h^{(L)} + \Delta_h^{(L)}/2)$. The temperature dependence of the principal order parameter $\tilde{S}^{(L)} = \left\langle \frac{3}{2}(\mathbf{L} \cdot \tilde{\mathbf{n}}^{(L)})^2 - \frac{1}{2} \right\rangle$, obtained as the major eigenvalue of the ordering matrix, together with the respective biaxiality order parameter in the principal axis frame $\tilde{\Delta}^{(L)} = \left\langle (\mathbf{L} \cdot \mathbf{m})^2 - (\mathbf{L} \cdot \tilde{\mathbf{l}}^{(L)})^2 \right\rangle$ and the rotation angle $\tilde{\theta}^{(L)}$, are shown in figure 6. As expected, $\tilde{S}^{(L)}$ is larger than the calculated extrapolation of the nematic order parameter $S$ below the transition temperature (dotted line in figure 4) and grows monotonously with decreasing temperature. The angle $\tilde{\theta}^{(L)}$ increases rapidly below the transition and moderately at even lower temperatures. Notably, the biaxiality parameter $\tilde{\Delta}^{(L)}$ increases monotonously and is about an order of magnitude larger than $\Delta_h^{(L)}$ of figure 4. It should be stressed that none of the $\tilde{\mathbf{n}}^{(L)}, \tilde{\mathbf{l}}^{(L)}$ axes is a symmetry axis, as the ordering about either of them is highly polar (obviously, the order parameter $P = \langle \mathbf{y} \cdot \mathbf{m} \rangle$ remains invariant under the rotation about $\mathbf{m}$) and therefore, unlike $\mathbf{m}$, they cannot be termed "directors".

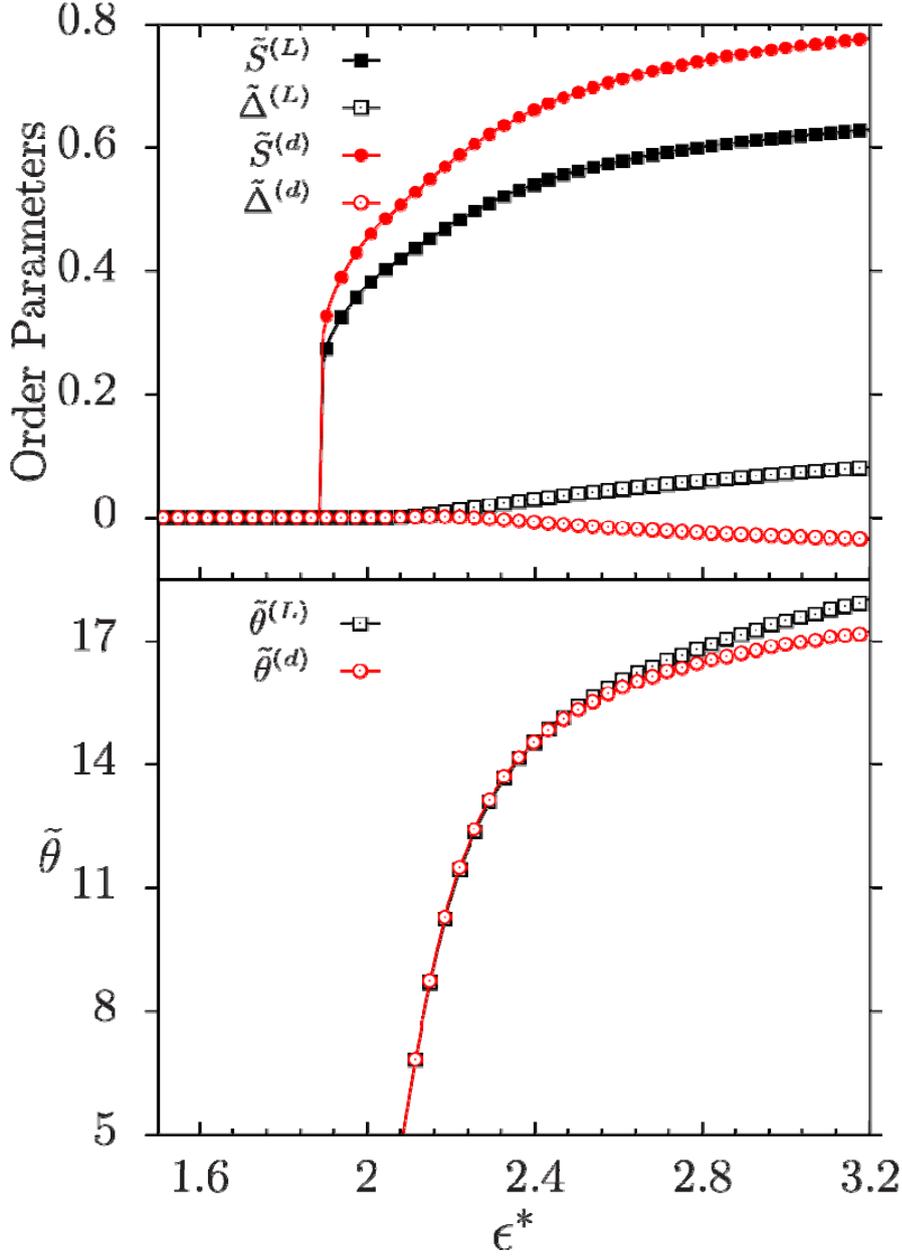

**Figure 6.** Temperature dependence of the principal order parameter $\tilde{S}^{(L)}$ of the mesogenic units, of the biaxiality order parameter $\tilde{\Delta}^{(L)}$ in the principal axis frame (top) and of the rotation angle $\tilde{\theta}^{(L)}$ of the later frame relative to the helix axis frame (bottom). The plots are calculated from the values of the order parameters $S_h^{(L)}$, $\Delta_h^{(L)}$ and $\delta_h^{(L)}$ in figure 4. Shown on the same diagrams are the principal order parameters $\tilde{S}^{(d)}$, $\tilde{\Delta}^{(d)}$ and the rotation angle $\tilde{\theta}^{(d)}$ associated with the spacer segment **d**.

Following the twisting of the director **m**, the principal axes $\tilde{\mathbf{n}}^{(L)}, \tilde{\mathbf{l}}^{(L)}$ also twist on moving along the helix axis $\mathbf{n}_h \| Z$, maintaining constant "cone angles" $\tilde{\theta}^{(L)}$ and $\tilde{\theta}^{(L)} + \pi/2$, respectively, with the helix axis. The tilted twisting of $\tilde{\mathbf{n}}^{(L)}$ bears a resemblance to the heliconical arrangement of the nematic director $\mathbf{n}$ in the twist-bend model, $N_{TB}$, of the nematic phase.[8,18,25] The resemblance is only superficial because, unlike the nematic director $\mathbf{n}$, the principal axis $\tilde{\mathbf{n}}^{(L)}$ is not a symmetry axis, global or local. Nevertheless, to pursue this resemblance a bit further, we disregard for the moment the fact that $\tilde{\mathbf{n}}^{(L)}$ and $\tilde{\mathbf{l}}^{(L)}$ are not directors and define for these principal axes the analogue of a "bend vector", as $\tilde{\mathbf{b}}^{(L)} = \tilde{\mathbf{n}}^{(L)} \times \nabla \times \tilde{\mathbf{n}}^{(L)} = -\tilde{\mathbf{l}}^{(L)} \times \nabla \times \tilde{\mathbf{l}}^{(L)}$. We also define the tilt pseudovector of these axes as $\tilde{\boldsymbol{\xi}}^{(L)} = (\tilde{\mathbf{n}}^{(L)} \cdot \mathbf{Z})(\mathbf{Z} \times \tilde{\mathbf{n}}^{(L)}) = -(\tilde{\mathbf{l}}^{(L)} \cdot \mathbf{Z})(\mathbf{Z} \times \tilde{\mathbf{l}}^{(L)})$ and the analogues of the "twist pseudoscalar", $t_{\tilde{\mathbf{n}}(L)} = \tilde{\mathbf{n}}^{(L)} \cdot (\nabla \times \tilde{\mathbf{n}}^{(L)})$ and $t_{\tilde{\mathbf{l}}(L)} = \tilde{\mathbf{l}}^{(L)} \cdot (\nabla \times \tilde{\mathbf{l}}^{(L)})$. Then, noting that the twist pseudoscalar for the $\mathbf{m}$ director is simply $-k$, i.e. $t_m = \mathbf{m} \cdot (\nabla \times \mathbf{m}) = -k$, and that $\tilde{\mathbf{b}}^{(L)}$ and $\tilde{\boldsymbol{\xi}}^{(L)}$ are in the direction of $\mathbf{m}$, we obtain the relations:

$$\tilde{\mathbf{b}}^{(L)} = -t_m \tilde{\boldsymbol{\xi}}^{(L)}, \quad t_{\tilde{\mathbf{n}}(L)} + t_{\tilde{\mathbf{l}}(L)} = t_m, \quad t_{\tilde{\mathbf{n}}(L)} t_{\tilde{\mathbf{l}}(L)} = t_m^2 \left( \tilde{\boldsymbol{\xi}}^{(L)} \right)^2 \tag{13}$$

which relate the bend vectors and the twist pseudoscalars of $\tilde{\mathbf{n}}^{(L)}$ and $\tilde{\mathbf{l}}^{(L)}$ to the twist of the $\mathbf{m}$ director and the respective tilts. These are simply geometrical relations, reflecting the generation of the twisting and bending of the $\tilde{\mathbf{n}}^{(L)}$ and $\tilde{\mathbf{l}}^{(L)}$ principal axes from the pure twisting of the director $\mathbf{m}$, and the "tilting" of these axes. In the present model, this tilting is a consequence of the loss of rotational symmetry about the nematic director as the transverse polar ordering sets in. It is important to note here that, as with any nematic phase of monoclinic symmetry,[34] the bend vector $\tilde{\mathbf{b}}^{(L)}$, the tilt pseudovector $\tilde{\boldsymbol{\xi}}^{(L)}$ and the twist pseudoscalars $t_{\tilde{\mathbf{n}}(L)}$, $t_{\tilde{\mathbf{l}}(L)}$ are, in principle, not unique attributes of the phase as whole but refer to particular molecular segments. Thus in the $N_X$ phase of flexible molecules consisting of several orientationally inequivalent groups of segments (for example, not identical mesogenic units, flexible spacer segments), there is a different set of $\tilde{\mathbf{b}}$, $\tilde{\boldsymbol{\xi}}$, $t_{\tilde{\mathbf{n}}}$, $t_{\tilde{\mathbf{l}}}$ for each such group. As an illustration, the three-segment toy dimer of figure 1, consisting of the two identical mesogenic segments $\mathbf{L}, \mathbf{L}'$ and a "spacer" $\mathbf{d}$, has, in analogy with the order parameters of the mesogenic segments, a set of parameters associated with the ordering of the $\mathbf{d}$ segment. The respective principal values of the order parameter $\tilde{S}^{(d)}$, the biaxiality $\tilde{\Delta}^{(d)}$ and rotation angle $\tilde{\theta}^{(d)}$ connecting the helix axis $\mathbf{n}_h$ to the principal axis $\tilde{\mathbf{n}}^{(d)}$ of the ordering tensor of the **d** segment, are shown in the temperature dependence graphs of figure 6 . Clearly, the "tilt" angle $\tilde{\theta}^{(d)}$ differs from $\tilde{\theta}^{(L)}$, although in this case the difference is small, due to the near rigidity of the toy-dimer (only $\pm\alpha$ torsions allowed) and the limited phase biaxiality. However, an immediate implication of $\tilde{\theta}^{(L)} \neq \tilde{\theta}^{(d)}$ is that the respective bend vector $\tilde{\mathbf{b}}^{(d)}$, the tilt pseudovector $\tilde{\boldsymbol{\xi}}^{(d)}$ and the twist pseudoscalars $t_{\tilde{\mathbf{n}}(d)}$, $t_{\tilde{\mathbf{l}}(d)}$ will differ from those of the $\mathbf{L}, \mathbf{L}'$ segments. Of course, $\tilde{\mathbf{b}}^{(d)}$, $\tilde{\boldsymbol{\xi}}^{(d)}$ $t_{\tilde{\mathbf{n}}(d)}$, $t_{\tilde{\mathbf{l}}(d)}$ satisfy the analogues of eq(13), relating them to the director twist $t_m$, which is the only measure of deformation that refers to the phase as a whole and not to particular segments. This is an essential distinction between the $N_X$ phase and the uniaxial nematic phase N, wherein the bend, splay and twist deformations refer to the nematic director $\mathbf{n}$ and are therefore segment-independent. This distinction applies also to the relation between phase polarity and deformations: In the N phase, the flexoelectric coefficients are phase properties relating the appearance of polar order to the vector deformations of the director $\mathbf{n}$. If these relations are carried over directly to the $N_X$ phase, the flexoelectric coefficients become segment-dependent. For example, establishing a formal relation between the bend vector $\tilde{\mathbf{b}}^{(L)}$ and the polarization vector $\mathbf{P}$ of the phase according to $\mathbf{P} = \tilde{e}_b^{(L)} \tilde{\mathbf{b}}^{(L)}$ would imply a bend flexoelectric coefficient $\tilde{e}_b^{(L)}$ that is different from the flexoelectric coefficient $\tilde{e}_b^{(d)}$ relating the **d**-segment bend vector $\tilde{\mathbf{b}}^{(d)}$ to $\mathbf{P}$. Similarly for the piezoelectric coefficients, relating the polarisation to the tilt vector. For example, $\mathbf{P} = \tilde{e}_p^{(L)} \tilde{\boldsymbol{\xi}}^{(L)}$ and the analogous expression involving $\tilde{\boldsymbol{\xi}}^{(d)}$ would imply different piezoelectric coefficients $\tilde{e}_p^{(L)} \neq \tilde{e}_p^{(d)}$. However, within the present model, the ratios of piezoelectric to flexoelectric coefficients are segment-independent and equal to the twist pseudoscalar of the phase, i.e. $\tilde{e}_p^{(L)} / \tilde{e}_b^{(L)} = \tilde{e}_p^{(L)} / \tilde{e}_b^{(L)} = -t_m = k$, which merely expresses the appearance of the bend as a combined result of the primary twisting of the **m** director and the tilting of the respective principal axis, both of which originate from the transverse polar ordering along the local $C_2$ axis. Stated briefly, the twisting and bending of the various principal axes ($\tilde{\mathbf{n}}^{(L)}$, $\tilde{\mathbf{n}}^{(d)}$, etc.) that can be defined for segmental ordering tensors, and the associated heliconical configurations, fall out naturally from the model as by-products of the onset of polar ordering along the local $C_2$ axis. The bending vectors, tilts, and twists are in general segment-dependent and, as such, they are not unique characteristics of the order deformations in the $N_X$ phase. However, they all derive from the fundamental deformation of the phase, which is the pure twisting (no bend) of the director **m**.

In contrast to the above picture, the twist-bend deformation of the director **n** constitutes, according to the $N_{TB}$ model, the fundamental distinguishing feature of the phase, and is a consequence of the destabilisation of the uniaxial nematic phase as the bend elastic constant tends to vanishing or even negative values.[25] Polar ordering is not involved in this mechanism, and seems to have been completely ignored in the original development of the $N_{TB}$ model,[25] although a weak local polarity is necessarily present as a result of the bend deformation of **n**. To become applicable to the low temperature nematic phase of bimesogens, the original $N_{TB}$ model has been modified and extended in some ways. For one, a spontaneous twist-bend resulting from instabilities associated with the elastic constants would be expected to be of much larger, and strongly temperature dependent, pitch than actually observed for bimesogen systems. A recent attempt[21] to reconcile the original $N_{TB}$ model with such observations, and to introduce polar ordering, starts out with symmetric rigid bent-core molecules and a locally uniaxial and apolar mean field whose director **n** is pre-deformed in a heliconical configuration. The pitch and conical angle of this configuration are determined by requiring the pre-deformed **n** to align optimally with both arms of the bend molecule. The bend vector of the so imposed deformation becomes an axis of polar ordering and, of course, the pitch of the imposed helical deformation comes out to be of a few molecular lengths and to be essentially temperature independent. Not surprisingly, the bend elastic constant is found to decrease with decreasing temperature. Obviously this modelling moves in the reverse direction from our fully molecular formulation, wherein the tilted twisting emerges simply through the diagonalisation of an ordering tensor in an intrinsically polar and twisted medium, as opposed to being a forced deformation of an ad-hoc director field on which all the symmetry and ordering properties of the phase are then built. Moreover, as shown above, the heliconical configuration of some entity which is not endowed with the symmetry properties the nematic director **n**, does not necessarily imply a twist-bend deformation. A further, well-known, example is the heliconical configuration in the chiral SmC*, which has the same symmetry, albeit in a layered structure and with much larger pitch than the bimesogen $N_X$ phase, but is of course not termed a "twist-bend" phase.

Regarding the bend elastic constant, our results are in closer agreement with the phenomenological theory of ref[20], wherein the tendency towards transverse polar ordering renormalizes the effective elastic constant of the nematic phase through the bend flexoelectric coefficient. Of course, the description of the deformations in the $N_X$ phase in terms of the three elastic constants of the nematic phase is not meaningful as neither $\tilde{\mathbf{n}}^{(L)}$ nor $\mathbf{n}_h$ have the symmetries of the director $\mathbf{n}$ of the uniaxial apolar nematic phase. It was also shown recently[35] that the key assumption of the $N_{TB}$ model, i.e. a negative bend elastic constant, is not necessary for the formation of a heliconical configuration.

Aside from the origin of the helical deformations, and the entities undergoing the deformation, our molecular model and the $N_{TB}$ model differ fundamentally in the description of the orientational ordering of the molecules and this has direct implications on the interpretation of the related experimental measurements. Thus, based on the assumption of a uniaxial, apolar nematic director under a twist-bend deformation[9,10,18,19,25] the respective expressions for the spectral frequencies in eq. (12) would be[13]:

$$\left[\Delta \nu_Q\right]_{H \parallel Z} \sim S\left(3\cos^2\theta_0 - 1\right)/2 \quad ; \quad \left[\Delta \nu_Q\right]_{H \parallel Y} \sim S\left(3(\sin\theta_0 \sin\varphi(Z))^2 - 1\right)/2 \quad (14)$$

where $S$ stands for the order parameter of the mesogenic segments in the frame of nematic director $\mathbf{n}$, and $\theta_0$ is the constant tilt angle of the heliconical twist-bend deformation of $\mathbf{n}$. While the left hand expression is the equivalent to of the respective eq. (12), the expression for the "$\pi/2$-flipped" spectra is completely different form the one in eq. (12) and implies substantial variation of the spectra with the position $Z$ along the helix, therefore broadened line-shapes, in sharp contrast with the measurement.[13] To remedy this discrepancy of the $N_{TB}$ interpretation, it has been claimed[19] that the translational diffusion of the dimer molecules is fast enough (within the chiral domains but not across them) to average out the $Z$ dependence in eq. (14). Of course, no assumption of such selectively fast diffusion is necessary for the interpretation of the spectra according to eq(12).

## 5. Conclusions.

Based on a simplified molecular model of flexible achiral mesogenic dimers and on standard statistical mechanics approximations for the formulation of the free energy in nematics, we have developed a fully molecular theory for the low temperature nematic phase, $N_X$, exhibited by certain odd-spacer dimer liquid crystals.

The transition from the uniaxial nematic N to the $N_X$ phase is found to be driven by the onset of transverse polar ordering of the molecules, dictated by the packing of their intrinsically polar shapes. The $N_X$ phase has local monoclinc symmetry, with its single director $\mathbf{m}$ defining the direction of polar ordering. The spontaneous symmetry breaking of phase apolarity results in the helical twisting of the director $\mathbf{m}$ about a perpendicular direction of macroscopically full rotational symmetry. The pitch of the twisting is a few molecular lengths and both senses of twisting are thermodynamically equivalent, allowing for the formation of domains of opposite chirality.

The local biaxiality of the phase and the chiral shifting of the molecular conformations are found to be rather small. The combination of the pure twist of the director $\mathbf{m}$ with the monoclinic local symmetry is shown to produce a tilted-twisted (heliconical) configuration of the principal axes of the ordering tensors of the molecular segments and, more generally, of any tensor property. The pitch of the heliconical configurations is determined by the pitch of the pure twist of the director, $\mathbf{m}$, whilst the conical angle is not unique and varies with the molecular segment or tensor property considered. This is contrasted with the twist-bend model, $N_{TB}$, picture wherein the director itself assumes a heliconical configuration, which is therefore common to all molecular segments and tensor properties.

The results of the molecular theory are found to account successfully for key experimental observations on the $N_X$ phase, to call into question the proposal that the N-$N_X$ transition is driven by a negative bend elastic constant and to provide consistent interpretations on certain controversial points regarding the $N_X$ phase structure, symmetries and molecular ordering.

**Acknowledgments.**

This research has been supported through the Greek (NSRF) program THALES: NANOLICR, co-financed by the European Union (European Social Fund – ESF) and Greek national funds.